# Criteria and Tips for Choosing Suitable Journal for Your Paper

Mohamed Hammad[1,2*]

[1]EIAS Data Science Lab, College of Computer and Information Sciences, Prince Sultan University, Riyadh 11586, Saudi Arabia

[2]Information Technology Department, Faculty of Computers and Information, Menoufia University, Egypt

*Corresponding:
E-mail: mohammed.adel@ci.menofia.edu.eg; mohamedadelhaamad@gmail.com

## Abstract

Selecting the right journal for your research paper is a pivotal decision in the academic publishing journey. This paper aims to guide researchers through the process of choosing a suitable journal for their work by discussing key criteria and offering practical tips. The significance of this decision lies in the potential impact of your research, the alignment of author's findings with the journal's scope, and the reception your work receives in the scientific community. In this paper, we delve into the importance of meticulously matching your paper's subject matter with the journal's scope, assessing the journal's reputation and impact factor, understanding the intricacies of the submission process, and upholding ethical and open-access considerations. By following these guidelines, researchers can make informed decisions that increase the chances of their work being published in a respected journal that truly complements their research objectives and aspirations.

Keywords: Choosing a journal; Research paper; Submission process; Scope of the journal; Research impact.

## 1. Introduction

The rejection of high-impact work can occur when the chosen search topic does not align with the scope of the journal. This error can result in the squandering of valuable resources such as time, financial investments, and personal drive. One of the most frequently seen errors made by authors is the submission of a manuscript to an unsuitable journal. This mistake is prevalent among authors of varying levels of experience, including both novices and experts. The process of selecting an appropriate journal for one's submission might induce a considerable amount of stress. Ultimately, opting to submit one's work to a suitable academic journal will enhance the likelihood of achieving publication [1]. The selection process involves careful consideration of various factors to ensure that your work reaches the right audience and has a significant impact in your field such as:

- Should the author solely consider the impact factor of a journal?
- Should the author prioritize selecting a journal with a shorter turnaround time?
- What are the steps involved in journal publication?
- Would it not be advantageous to own a concise reference document for the purpose of selecting a magazine for consideration?

This paper addresses the pivotal question: How do researchers navigate the labyrinth of academic publishing to select the most fitting journal for their research? To answer this question, we will explore a set of criteria and provide valuable tips that will enable researchers to make informed choices. These criteria encompass assessing the alignment of the final paper with the journal's scope, evaluating the journal's reputation and impact factor, understanding the intricacies of the submission process, and upholding ethical standards and open-access considerations. In a world where the pursuit of knowledge knows no bounds, the choice of a journal must be a well-informed decision, one that reflects your research's importance, the relevance of your findings to a specific field, and your career aspirations. Through this paper, we aim to equip researchers with the knowledge and insights necessary to make this critical decision, one that will ultimately shape the trajectory of their academic and scientific journey.

Before choosing the right journals, important questions about publishing in scientific journals need to be addressed first to ensure the optimal dissemination of your research. These questions are as follows:

- What scientific journals are approved for publication?
- What is Scopus or ISI Web of Science?
- How to know if this journal is in Scopus or not?
- How to know if this journal is in Web of Science or not?
- What is the Impact factor (IF)?
- How much is the impact factor? what is better High or Low?
- How to check impact factor of Scopus indexed journals?
- What are the most important publishers that contain scientific journals with an impact factor and be free to publish?
- What is the difference between open access and subscription journals?

The answers to these previous questions are discussed in the next Section. After that Section 3 shows how to prepare your paper for submission. Section 4 shows how to choose the best journal for your paper. Section

5 provides a checklist for selecting the best journal. Section 6 shows the most serious issues that faced the authors in publishing the paper. Finally, Section 7 concludes all the information in this paper.

## 2. Important questions about publishing in scientific journals

Publishing scientific research in reputable journals is a crucial step in the academic and scientific community. As researchers aim to disseminate their findings and contribute to the body of knowledge, they often encounter various questions and considerations in the process. Here, we address some of the most important questions that researchers should contemplate when navigating the world of scientific publishing:

1) What scientific journals are approved for publication?

The process of selecting a scientific journal for publication involves several crucial considerations. Researchers should aim to publish their work in reputable journals that adhere to rigorous standards. Approval for publication depends on the alignment of the research with the journal's scope, the quality of the research, and its adherence to ethical guidelines. Generally, scientific journals are approved by Scopus or ISI Web of Science [2].

2) What is Scopus and ISI Web of Science?

Scopus [3] and ISI Web of Science [4] are two widely recognized bibliographic databases that index academic journals, conference proceedings, and other scholarly publications. They serve as invaluable resources for researchers, helping them identify relevant literature and evaluate the impact and quality of journals. Table 1 shows differences between Scopus and Web of Science.

Table 1. Scopus Versus Web of Science

| Aspect | Scopus | Web of Science |
|---|---|---|
| Publisher | Elsevier | Clarivate Analytics |
| Coverage | Multidisciplinary; strong in science, technology, and medicine | Multidisciplinary; covers social sciences and arts and humanities too |
| Citation Index | Scopus Citation Index | Web of Science Citation Index |
| Journal Inclusion | Active inclusion with regular updates | Selective inclusion, rigorous selection process |
| Coverage of Conferences | Extensive coverage of conference proceedings | Limited coverage of conference proceedings |
| Author Profiles | Author profiles and affiliations available | Author profiles and affiliations available |
| Cited Reference Searching | Yes, allows cited reference searching | Yes, allows cited reference searching |
| Journal Impact Factor | CiteScore is used as an alternative to Impact Factor | Journal Impact Factor is widely recognized |
| Collaboration with Publishers | Direct collaboration with publishers for data integration | Content is gathered from published materials |
| Open Access Journals | Includes many open access journals | Includes open access journals but with a more limited selection |
| Database Size | Large database with extensive coverage | Comprehensive, with a smaller but highly influential selection |
| Funding | Subscription-based, requires access through institutions | Subscription-based, requires institutional access |
| User-Friendly Interface | User-friendly and intuitive interface | User-friendly interface with various search options |

3) How to know if this journal is in Scopus or not?

To ascertain whether a journal is indexed in Scopus, researchers can visit the Scopus website and use the journal search feature [3]. A journal's inclusion in Scopus indicates that it has met certain quality and editorial criteria. Figure 1 shows a screenshot of a journal that appears to be in Scopus. This Figure shows that the journal "Journal of Business Research" is in Scopus with CiteScore of 16.

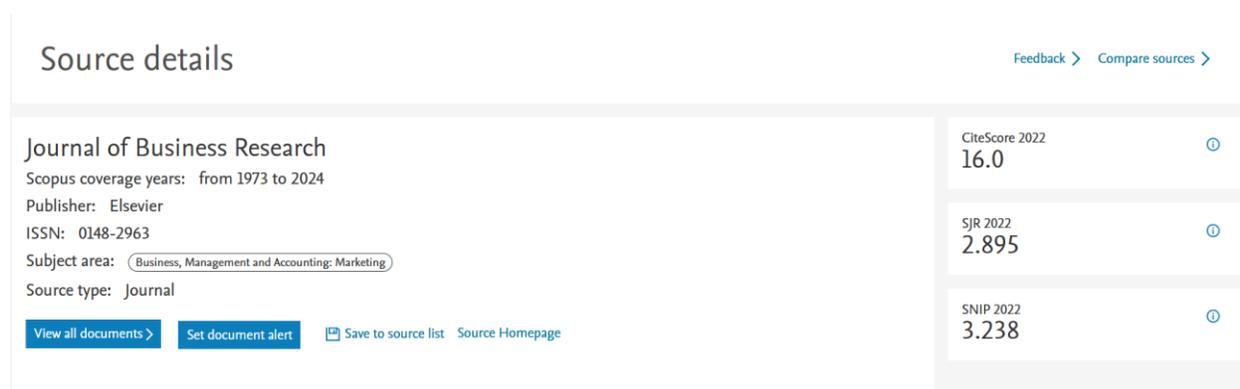

Figure 1. Screenshot of Scopus results on Journal of Business Research

4) How to know if this journal is in Web of Science or not?

To determine if a journal is indexed in the Web of Science, researchers can access the Web of Science database or the Master Journal List available on the Clarivate Analytics website [4]. Inclusion in Web of Science reflects a journal's international recognition and significance. Figure 2 shows a screenshot of a journal that appears to be in the Web of Science. This Figure shows that the journal "Journal of Business Research" is in Web of Science with Impact Factor of 11.3.

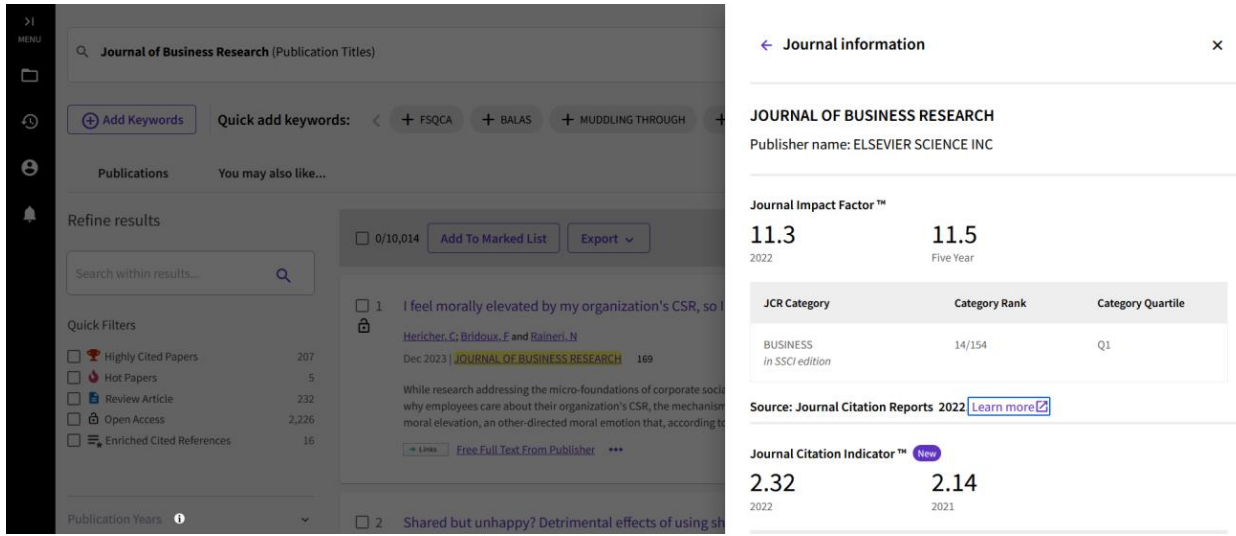

Figure 2. Screenshot of Web of Science results on Journal of Business Research

5) What is the Impact factor (IF)?

The evaluation of a journal is contingent upon the quantity of research articles published inside the journal, as well as the extent to which the research published within the journal is utilized as sources in other journals with IF [5]. It is used as an indicator of a journal's influence and prestige. It is calculated by Clarivate Analytics, and it's primarily used in the context of the Web of Science database. IF computed as the following:

$$IF = \frac{(Total\ Citations\ in\ Year\ 1 + Total\ Cittions\ in\ Year\ 2)}{(Total\ Citable\ Articles\ in\ Year\ 1 + Total\ Citable\ Articles\ in\ Year\ 2\ )} \quad (1)$$

For example, if a journal had 500 citations in 2022 and 600 citations in 2023, and it published 200 citable articles in 2022 and 220 in 2023, the calculation would be:

IF = (500 + 600) / (200 + 220) = 1100 / 420 = 2.62

6) How much is the impact factor? what is better High or Low?

Impact Factors can vary widely, with some journals having high values and others having lower values. The "better" value depends on your goals. A high Impact Factor may signify a journal's prominence but also potentially less accessibility. A lower Impact Factor may indicate a more specialized readership.

7) How to check impact factor of Scopus indexed journals?

Scopus does not officially calculate or provide the IF for journals like Clarivate Analytics does for journals indexed in the Web of Science database [6]. The IF, as calculated by Clarivate Analytics, is specific to the Web of Science database and is based on their citation data. However, Scopus offers its own set of journal metrics, such as the CiteScore and Scimago Journal Rank (SJR), which can provide insight into the influence and significance of journals indexed in Scopus. These metrics are calculated by Scopus itself and can be used to evaluate the impact of journals. CiteScore is similar to the IF and is calculated by dividing the number of citations received by a journal in a specific year by the total number of documents published by the journal in the three previous years. It's updated annually. SJR is another metric that takes into account both the number of citations received and the quality of the citing sources. It's designed to give more weight to citations from highly ranked journals.

8) What are the most important publishers that contain scientific journals with an impact factor and be free to publish?

Several well-known publishers offer journals with Impact Factors and provide open access options. Some notable publishers in this category include Dutch Publisher Elsevier [7], American Publisher IEEE [8], German Publisher Springer [9], American Publisher Wiley [10], British Publisher Taylor & Francis [11], British Publisher Nature [9], Public Library of Science (PLOS) [12] and Frontiers [13]. These publishers offer researchers the opportunity to publish their work with high visibility and accessibility.

9) What is the difference between open access journals and subscription journals?

Open access journals make research freely available to the public, while subscription journals require payment or access through institutional subscriptions [14]. Researchers should consider factors like funding availability, target audience, and licensing options when deciding between the two. Open access can increase accessibility but may involve publication fees. Table 2 shows the key differences between the two types.

Table 2. Comparison of Open Access and Subscription Journals

| Aspect | Open Access | Subscription |
| --- | --- | --- |
| Access to Content | Freely accessible to the public without access restrictions. | Requires a subscription or purchase for access. |
| Business Model | May charge publication fees (APCs) or use alternative funding models. | Rely on subscription revenue advertising, and reprints. |
| Copyright and Licensing | Often use open licenses (e.g., CC-BY) allowing reuse with attribution. | Copyright retained by the journal; access rights more restricted. |
| Peer Review | Can have rigorous peer review processes | Can have rigorous peer review processes |
| Impact and Prestige | Not determined by access model; can have high-impact publications. | Not determined by access model; can have high-impact publications. |
| Funding and Sustainability | Rely on publication fees, subsidies, grants, or institutional support. | Rely on subscription revenue for sustainability. |

## 3. Prepare for submitting

Preparing a manuscript for submission to a journal is a critical step in the research and publication process. To increase the chances of acceptance, it's important to meet certain criteria and follow specific tips. Here are the criteria and tips for preparing a manuscript for submission:

## 3.1. Criteria (1): Ensure that there is alignment between the topic matter of your paper and the stated objective and scope of the publication

When writing a manuscript for submission to an academic journal, it is crucial to ensure that the subject matter of your article aligns with the objective and scope of the publication. This alignment serves as a primary criterion for consideration. One of the most prevalent and preventable causes for the rejection of a journal submission is a lack of alignment between the content of the manuscript and the specific objectives and scope of the journal. Typically, this information can be easily accessed on the homepage of the publication. Please locate a section inside the document labeled "About the Journal," "Full Aims and Scope," or a similar heading. By perusing this webpage, one can acquire essential insights regarding the compatibility of their research with the publication in question. Figure 3 shows a screenshot of the aim and scope from the home page of Journal of Business Research.

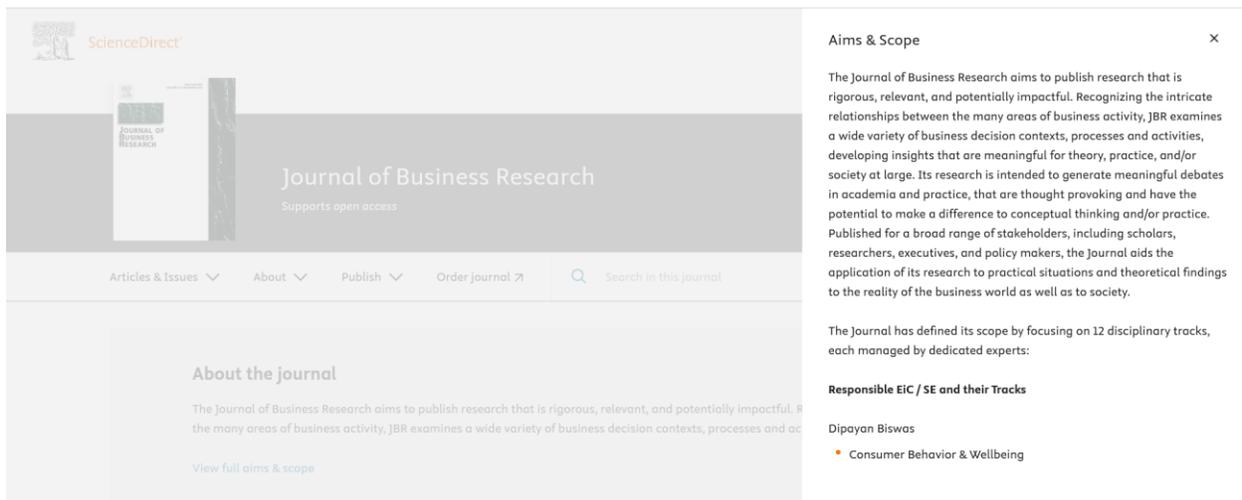

Figure 3. Screenshot of the aim and scope from the home page of Journal of Business Research

**Tip1**: Familiarize yourself with the stated objectives and coverage of the journal, followed by a thorough examination of the titles and abstracts of works featured in the journal's most recent publications. This will aid authors in refining the scope of issues that align with the journal's areas of interest.

## 3.2. Criteria (2): Journal's restrictions

- *Article Type*:

A publication that does not match your article's genre will likely reject it. As an illustration, some academic journals, including the British Journal of Surgery [15], do not publish case studies. Thus, the "Information for Authors" section of the intended journal must be checked for restrictions. Screenshot in Figure 4 illustrates the various types of papers that can be accepted by the journal Research Policy.

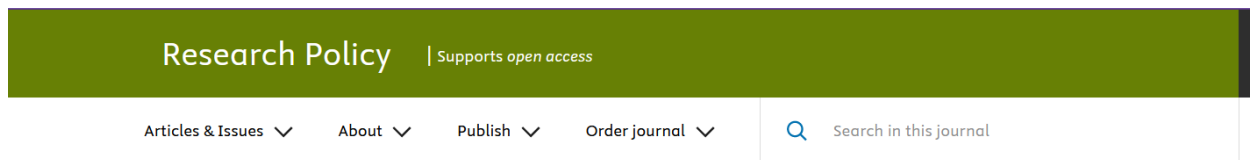

Figure 4. Screenshot shows the types of paper from Research Policy journal

- *Word Count*:

It is imperative to acknowledge the constraints associated with word limits. For instance, in the scenario where the manuscript consists of 7,000 words and the journal's stipulated word limit for papers is 4,000 words, it becomes evident that a significant rewrite will be necessary.

- *Publication Cost*:

The expense associated with publication can be perceived as a constraint, given that several journals impose substantial manuscript processing costs. Additional charges may be incurred for open access, exceeding a certain page limit, or including color figures.

3.3. Criteria (3): Impact Factor (IF)

The IF continues to serve as the primary metric for assessing the quality and prestige of a scholarly publication. While the allure of submitting a paper to a journal with a high IF may be strong, it is crucial to engage in an objective assessment of one's research to ascertain its suitability for publication in a prestigious journal. Alternatively, one may face the potential consequence of expending significant time and effort in the repetitive process of resubmitting and reformatting their paper for multiple academic journals.

### 3.4. Criteria (4): Turnaround time for articles submitted to the journal

**Definition** - Turnaround time (TAT) or Response time (RT) refers to the temporal duration between the moment a process is submitted and the moment it is fully completed.

- The timeliness of responses plays a crucial role in the submission process, particularly for authors. Authors tend to prioritize expeditious decision-making in order to achieve enhanced performance and swifter decision outcomes [16].
- Irrespective of the methodology employed for computation, the reduction of the turnaround time is evidently advantageous for authors and editors, particularly if it can be achieved without compromising the quality of the review process.
- Nevertheless, authors are not solely concerned with the duration required for obtaining external reviews. Some authors also seek information regarding the time it takes for a manuscript to be submitted, accepted, or rejected. This process differs significantly since it may entail multiple changes, during which the document remains under the author's control.
- There is a curiosity among individuals regarding the duration required for an officially accepted work to be published in both print and electronic formats.

**Tip2**: Ask colleagues about their experiences in specific journals. This will give the authors insider knowledge about the working on individual journals.

Authors can find the review time of some journals from the home page [17] of the journal select View all insights as shown in Figure 5 or find it directly on the home page of the journal as shown in Figure 5.

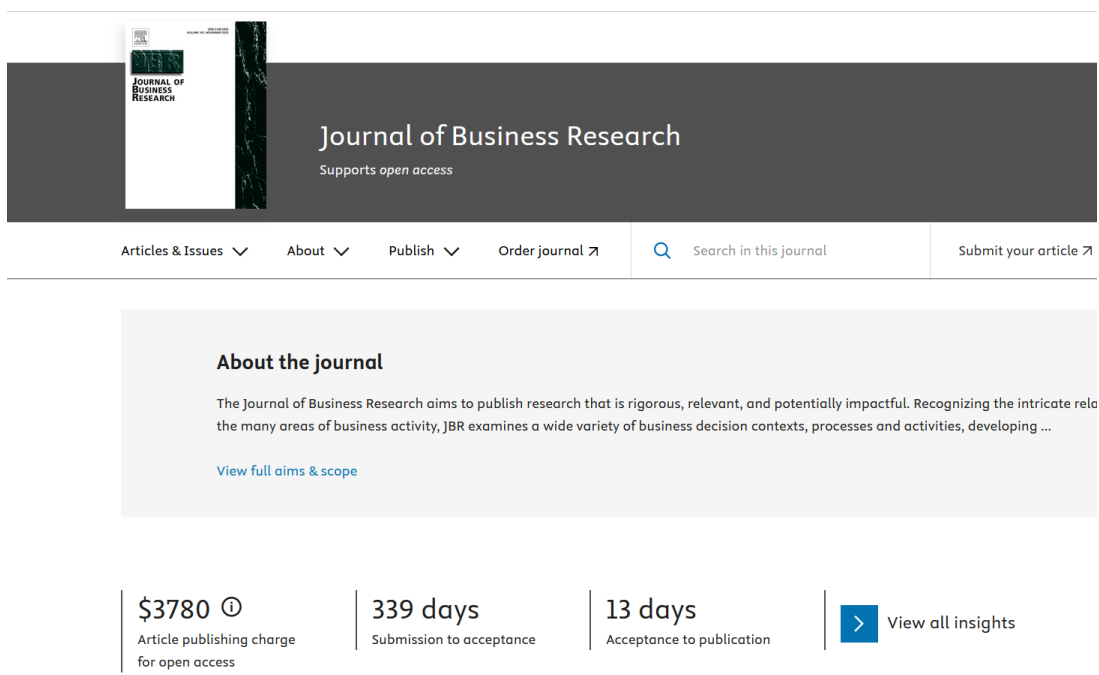

Figure 5. Publication timeline of Journal of Business Research form View all insights

### 3.5. Criteria (5): The readership and target audience

The choice between submitting to a multidisciplinary journal or a specialty journal is a critical decision [18,19]. The nature of the target journal significantly impacts the readership and audience for the research, and it has implications for how authors tailor their manuscript to meet the expectations of that readership. The distinction between multidisciplinary and specialty journals is particularly relevant in discussions about readership and target audience. Authors must consider how well their research aligns with the scope and focus of the chosen journal and, by extension, the nature of the readership they intend to reach.

- Multidisciplinary Journals

When authors opt for multidisciplinary journals, they are selecting a platform that caters to a diverse and potentially broad readership. The research published in these journals spans various disciplines, making them an ideal choice when the aim is to communicate research findings to a wide audience with different backgrounds and research interests. A manuscript prepared for submission to a multidisciplinary journal should be framed and written to be inclusive, ensuring accessibility and comprehensibility for a wide range of readers.

- Specialty Journals

On the other hand, specialty journals maintain a focused readership, consisting of experts, scholars, and professionals who possess in-depth knowledge and specific interests in the field or subfield. For authors targeting specialty journals, the manuscript must be finely tuned to cater to the highly specialized nature of the readership. The writing, methodology, and discussion should reflect a deep understanding of the subject matter and the expectations of the journal's expert audience.

Understanding the nature of the target journal and its readership is an integral part of preparing a manuscript for submission. Authors should consider not only the relevance of their research to the chosen journal but also how the manuscript is crafted to meet the expectations of the readers. This entails the use of appropriate terminology, framing, and depth of analysis.

Furthermore, the decision to submit to a multidisciplinary or specialty journal should be informed by the research's level of specialization and the degree of interdisciplinarity. If the work spans multiple fields or seeks to bridge gaps between disciplines, a multidisciplinary journal may be the optimal choice. Conversely, if the research is highly specialized and of interest primarily to experts in a specific area, then a specialty journal may be the most suitable venue.

### 3.6. Criteria (6): Journal visibility

After the publication of the work, it should be readily accessible and discoverable for fellow scholars. The prominence of journals is a significant factor in this context.
- Is the journal indexed in electronic databases?
- Is the journal included in the ISI's Web of Science index?
- Is the publication included in widely recognized subject-specific databases within the author's field?

This practice enhances the discoverability of writers' research inside their specific domain and potentially augments the frequency of citations for their scholarly articles. Figure 6 illustrates the process of abstracting and indexing for the Journal of Business Research.

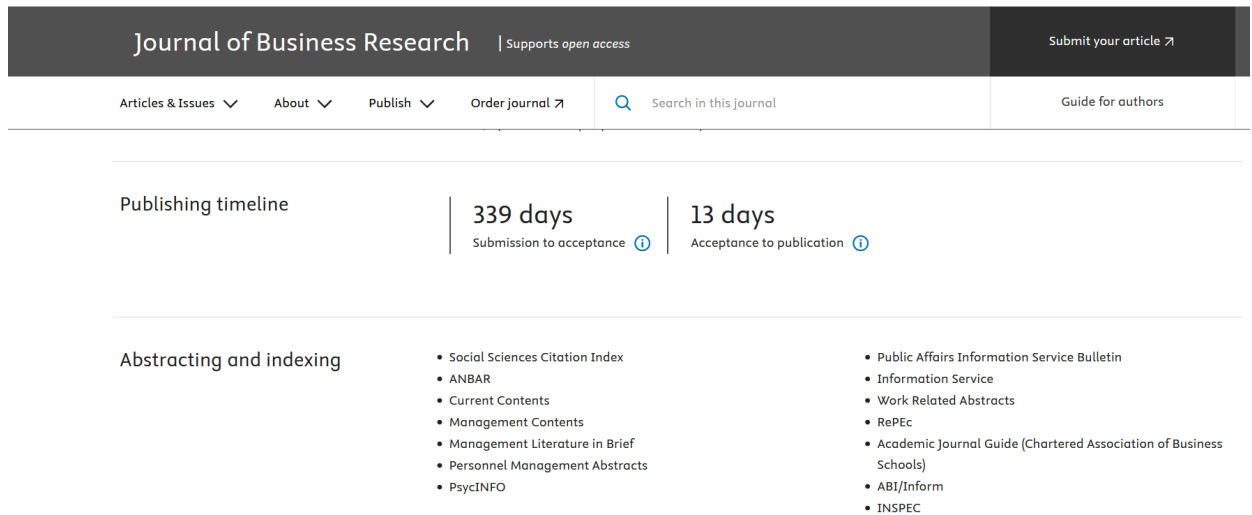

Figure 6. Screenshot for the abstracting and indexing of Journal of Business Research

**Tip3**: Conduct a rapid keyword search within the preferred database of the author. Does the journal appear as the first result in search queries?

4. What is the best journal for my paper?

Publication of a research paper poses a formidable challenge for authors, and this challenge is exacerbated when one considers the inherent risk of rejection associated with submitting a manuscript to a journal that may not be the most suitable fit. The selection of the right journal is a critical decision, with profound implications for the research's visibility and dissemination within the academic community. To navigate this complexity, some publisher introduced a tool to find the best journals of the submitted papers. The common tools are summarized as follows:

a) Elsevier's Journal Finder tool [20]: It offers a solution to this challenge, facilitating the process of identifying journals that are the best match for a particular manuscript. This tool assists authors in making well-informed decisions by streamlining the journal selection process. It takes into account a range of factors, such as the manuscript's abstract, keywords, and objectives, and matches these elements with the scopes and aims of journals in the Elsevier database. By doing so, it significantly reduces the likelihood of submitting a paper to a journal that is incongruent with its content. Figure 7 shows a screenshot of the journal finder tool page.

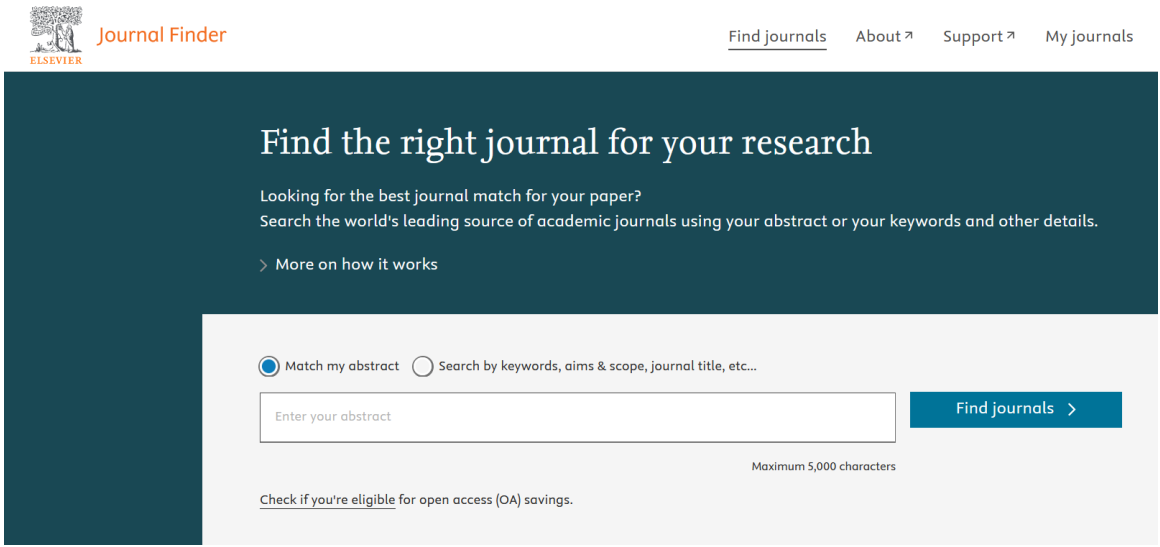

Figure 7. Screenshot of the Elsevier's journal finder tool page

b) Wiley's Journal Finder [21]: It serves as a valuable resource for authors seeking to identify journals that best suit their research. This tool streamlines the complex process of journal selection by leveraging technology to match the content and focus of a manuscript with the objectives and scope of journals within Wiley's extensive portfolio. Figure 8 shows a screenshot of Wiley's Journal Finder tool page.

Figure 8. Screenshot of the Wiley's Journal Finder page

c) Taylor & Francis Journal Suggester [22]: is a user-friendly tool designed to streamline the process of journal selection for authors. This tool employs technology to simplify and optimize the manuscript-to-journal matching process, providing a valuable service to academic authors. It leverages specific manuscript attributes, such as keywords, abstracts, and research objectives, to recommend a curated list of journals within the Taylor & Francis publishing portfolio that best aligns with the research's focus. Figure 9 shows a screenshot of the Taylor & Francis Journal Suggester tool page.

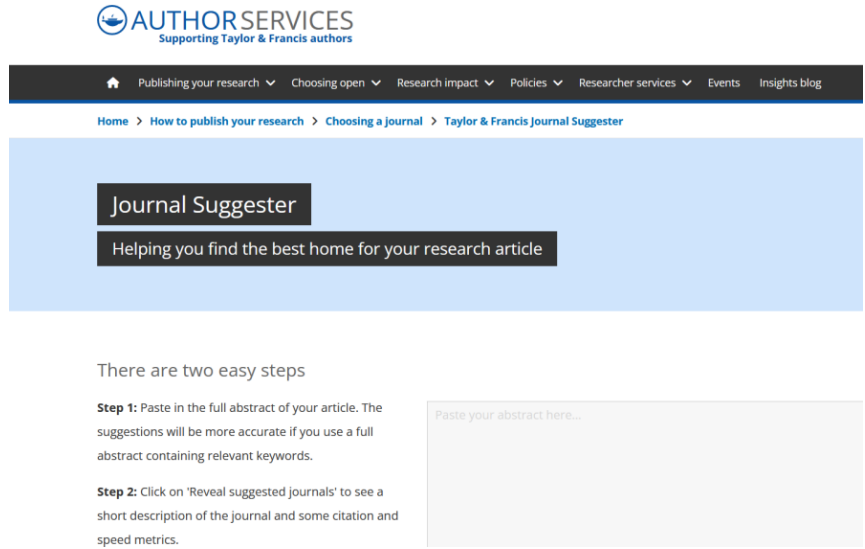

Figure 9. Screenshot of the Taylor & Francis Journal Suggester page

a) IEEE publications recommender [23]: It is a valuable tool designed to help authors identify the most suitable IEEE (Institute of Electrical and Electronics Engineers) publication for their research. It simplifies the complex process of journal selection by using advanced algorithms and data analysis. By considering key attributes of a manuscript, such as keywords, abstract, and research objectives, the tool generates a list of IEEE publications that closely match the manuscript's content, thereby facilitating informed decision-making. Figure 10 shows a screenshot of the IEEE publishing recommender tool page.

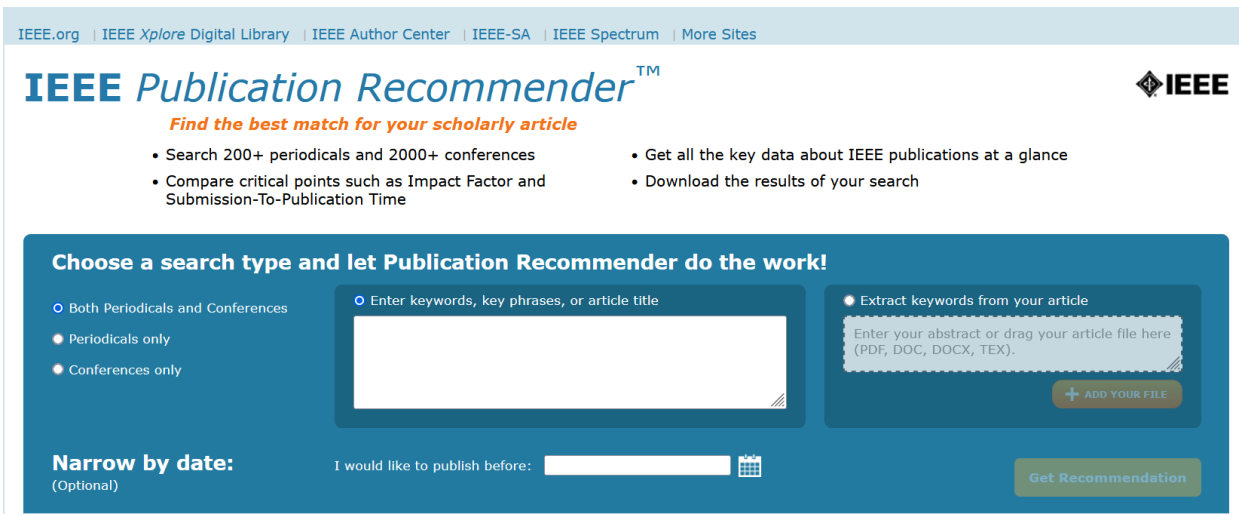

Figure 10. Screenshot of the IEEE publishing recommender page

## 5. Checklist for journal selection

Before submitting the paper, authors should ensure of all points in this checklist are in their paper:

1) Does the topic of your paper align with the thematic focus of the journal?
2) Is the intended article type compatible with the journal's submission criteria?
3) Does the journal cater to the intended readership of the authors? Do bibliographic and subject-specific databases list the journal?
4) Does the journal have an online version? Does the journal's impact factor meet authors' needs?
5) Is the journal known for its academic prestige among peers and colleagues?
6) How long does it take for journal papers to be processed and returned to authors?
7) What is the frequency of publication for the journal on an annual basis?
8) What are the fees associated with publishing?
9) Does the manuscript adhere to the journal's prescribed length and structural requirements.

## 6. The Most serious issues in publishing

Academic publishing is underpinned by the principles of integrity, honesty, and transparency. The scholarly community relies on the veracity of research findings and the attribution of credit to the rightful authors. However, the most serious issues in publishing are those related to fabrication, falsification, and plagiarism [24] (FFP), which compromise the integrity of the academic publishing process. Addressing these issues is paramount for upholding the credibility of scholarly work.

1. *Fabrication*:

Fabrication pertains to the intentional generation of data or outcomes that have not genuinely transpired. This is a severe breach of research ethics as it undermines the trustworthiness of academic publications. When fabricated data are included in research papers, it can lead to false conclusions, misinformed policies, and wasted resources. It erodes the foundation of knowledge on which academic research is built.

2. *Falsification*:

Falsification involves the manipulation or altering of data, results, or images to present a distorted or misleading account of the research. This distortion can manifest in various forms, such as selectively excluding data points, misrepresenting results, or altering visual representations (e.g., graphs and images). Falsification not only misleads readers but also undermines the robustness of scientific inquiry.

3. *Plagiarism*:

Plagiarism refers to the practice of utilizing another individual's concepts, language, or intellectual output without appropriate acknowledgment or attribution. It is a significant issue in publishing as it misappropriates the intellectual contributions of others and breaches the ethical foundations of scholarly communication. Plagiarism can take various forms, from verbatim copying of text to the uncredited use of ideas or concepts.

*4. Duplicate Submissions:*

Submitting the same manuscript for evaluation simultaneously to different journals is considered unacceptable. If the paper is published, it might be retracted on grounds of duplicate submission.

*5. Using the artificial intelligence (AI) software:*

AI technologies provide innovative research tools and enhance efficiency and accuracy in the research process. The most common AI techniques among these tools is ChatGPT [25], which is an advanced language model that has been developed by OpenAI. The model is built upon the GPT-3 framework and is specifically engineered to provide text answers that closely resemble human conversation. ChatGPT has the capability to be employed in a multitude of jobs related to comprehending and producing natural language. These tasks include but are not limited to responding to inquiries, participating in text-based dialogues, furnishing clarifications, and other similar endeavors. The technology exhibits versatility and finds utility in several domains such as chatbots, virtual assistants, customer service, content development, and other contexts that necessitate the usage of natural language processing for engaging in conversational interactions with users.

However, it is not recommended to rely entirely on these AI tools in research production, as most of these tools rely on language models, meaning that you are not dealing with a program that understands what it is writing [26]. These programs are trained to notice certain patterns in word order within a specific context and produce a final product based on this process. The tool may provide you with a great or very bad result depending on the context, and there are many dimensions on which the result provided by this tool depends, such as prompt engineering. In the end, as in the entire scientific process, you must be skeptical and critically evaluate everything you read and everything these tools give you. Most of the time, to get the best possible result from the tool you are using, you will fiddle a little with the settings or the prompt you gave to the tool until you reach a satisfactory result. My recommendation is that you must deal with artificial intelligence as your assistant who does not understand anything in your specialty to carry out boring and monotonous tasks and improve your productivity and not treat it as if it is an experienced scientist in the field.

## 7. Conclusion

The process of publishing your research paper in a suitable journal entails a comprehensive journey that begins with meticulous journal selection, considering factors such as scope, readership, and impact factor. Journal finder tools, such as Elsevier's Journal Finder, Wiley's Journal Finder, and IEEE Publications Recommender, can assist in this selection process, aiding authors in identifying journals that align with their research content. Ethical considerations, including the prevention of plagiarism, fabrication, and falsification, are paramount in maintaining the integrity of scholarly work. By carefully navigating these steps, authors can optimize the reach and impact of their research within the academic community.